\documentstyle[10pt,epsf,epsfig,hangcaption,xspace,amssymb,amsfonts,amsmath,amsthm,cite,
dp_delphititle,lineno,rotating]{dp_delphi}
\makeindex
\def\DpPaperGroup{EP}
\def\DpPaperRef{2002-089}
\def\DpDate{13 November 2002}
\def\DpAuthors{DELPHI Collaboration}
\def\DpSubmit{(Accepted by Phys. Lett. B)}
\def\DpTitle {{Inclusive b~Decays to\\
 Wrong Sign Charmed Mesons }}
\def\DpComment{ }
\def\DpEMail{ }

\begin{document}
\makeatletter
\newcount\@tempcntc
\def\@citex[#1]#2{\if@filesw\immediate\write\@auxout{\string\citation{#2}}\fi
  \@tempcnta\z@\@tempcntb\m@ne\def\@citea{}\@cite{\@for\@citeb:=#2\do
    {\@ifundefined
       {b@\@citeb}{\@citeo\@tempcntb\m@ne\@citea\def\@citea{,}{\bf ?}\@warning
       {Citation `\@citeb' on page \thepage \space undefined}}%
    {\setbox\z@\hbox{\global\@tempcntc0\csname b@\@citeb\endcsname\relax}%
     \ifnum\@tempcntc=\z@ \@citeo\@tempcntb\m@ne
       \@citea\def\@citea{,}\hbox{\csname b@\@citeb\endcsname}%
     \else
      \advance\@tempcntb\@ne
      \ifnum\@tempcntb=\@tempcntc
      \else\advance\@tempcntb\m@ne\@citeo
      \@tempcnta\@tempcntc\@tempcntb\@tempcntc\fi\fi}}\@citeo}{#1}}
\def\@citeo{\ifnum\@tempcnta>\@tempcntb\else\@citea\def\@citea{,}%
  \ifnum\@tempcnta=\@tempcntb\the\@tempcnta\else
   {\advance\@tempcnta\@ne\ifnum\@tempcnta=\@tempcntb \else \def\@citea{--}\fi
    \advance\@tempcnta\m@ne\the\@tempcnta\@citea\the\@tempcntb}\fi\fi}
 
\makeatother
\begin{titlepage}
\pagenumbering{roman}
\CERNpreprint{\DpPaperGroup}{\DpPaperRef} 
\date{{\small\DpDate}} 
\title{\DpTitle} 
\address{\DpAuthors} 
\begin{shortabs} 
\noindent
The production of wrong sign charmed mesons~$b\rightarrow\bar
D_{(s)}X$, $D_{(s)}=(D^0,D^+,D_s)$, is studied using the data
collected by the DELPHI experiment in the years 1994 and 1995.

Charmed mesons in $Z\rightarrow b\bar b$~events are exclusively
reconstructed by searching for the decays~$D^0\rightarrow K^-\pi^+$,
$D^+\rightarrow K^-\pi^+\pi^+$ and
$D_s^+\rightarrow\phi\pi^+\rightarrow K^+K^-\pi^+$. The wrong sign
contribution is extracted by using two discriminant variables: the
charge of the $b$-quark at decay time, estimated from the charges of
identified particles, and the momentum of the charmed meson in the
rest frame of the $b$-hadron.

The inclusive branching fractions of $b$-hadrons into wrong sign charm
mesons are measured to be:
\begin{eqnarray*}
 {\mathcal B}(b\rightarrow\bar D^0X)+{\mathcal B}(b\rightarrow D^-X) & = &
  (9.3\pm 1.7(stat)\pm 1.3(syst)\pm 0.4({\mathcal B}))\%~, \\
 {\mathcal B}(b\rightarrow D_s^-X) & = & 
  (10.1\pm1.0(stat)\pm 0.6(syst)\pm 2.8({\mathcal B}))\%
\end{eqnarray*}
where the first error is statistical, the second and third errors are
systematic.
\end{shortabs}
\vfill
\begin{center}
\DpSubmit \ \\ 
\DpComment \ \\
\DpEMail \ \\
\end{center}
\vfill
\clearpage
\headsep 10.0pt
\addtolength{\textheight}{10mm}
\addtolength{\footskip}{-5mm}
\begingroup
%
\newcommand{\DpName}[2]{\hbox{#1$^{\ref{#2}}$},\hfill}
\newcommand{\DpNameTwo}[3]{\hbox{#1$^{\ref{#2},\ref{#3}}$},\hfill}
\newcommand{\DpNameThree}[4]{\hbox{#1$^{\ref{#2},\ref{#3},\ref{#4}}$},\hfill}
\newskip\Bigfill \Bigfill = 0pt plus 1000fill
\newcommand{\DpNameLast}[2]{\hbox{#1$^{\ref{#2}}$}\hspace{\Bigfill}}
\small
\noindent
\DpName{J.Abdallah}{LPNHE}
\DpName{P.Abreu}{LIP}
\DpName{W.Adam}{VIENNA}
\DpName{P.Adzic}{DEMOKRITOS}
\DpName{T.Albrecht}{KARLSRUHE}
\DpName{T.Alderweireld}{AIM}
\DpName{R.Alemany-Fernandez}{CERN}
\DpName{T.Allmendinger}{KARLSRUHE}
\DpName{P.P.Allport}{LIVERPOOL}
\DpName{U.Amaldi}{MILANO2}
% ...

\normalsize
\endgroup
\titlefoot{Department of Physics and Astronomy, Iowa State
     University, Ames IA 50011-3160, USA
    \label{AMES}}
\titlefoot{Physics Department, Universiteit Antwerpen,
     Universiteitsplein 1, B-2610 Antwerpen, Belgium \\
     \indent~~and IIHE, ULB-VUB,
     Pleinlaan 2, B-1050 Brussels, Belgium \\
     \indent~~and Facult\'e des Sciences,
     Univ. de l'Etat Mons, Av. Maistriau 19, B-7000 Mons, Belgium
    \label{AIM}}
\titlefoot{Physics Laboratory, University of Athens, Solonos Str.
     104, GR-10680 Athens, Greece
    \label{ATHENS}}
\titlefoot{Department of Physics, University of Bergen,
     All\'egaten 55, NO-5007 Bergen, Norway
    \label{BERGEN}}
\titlefoot{Dipartimento di Fisica, Universit\`a di Bologna and INFN,
     Via Irnerio 46, IT-40126 Bologna, Italy
    \label{BOLOGNA}}
\titlefoot{Centro Brasileiro de Pesquisas F\'{\i}sicas, rua Xavier Sigaud 150,
     BR-22290 Rio de Janeiro, Brazil \\
     \indent~~and Depto. de F\'{\i}sica, Pont. Univ. Cat\'olica,
     C.P. 38071 BR-22453 Rio de Janeiro, Brazil \\
     \indent~~and Inst. de F\'{\i}sica, Univ. Estadual do Rio de Janeiro,
     rua S\~{a}o Francisco Xavier 524, Rio de Janeiro, Brazil
    \label{BRASIL}}
\titlefoot{Coll\`ege de France, Lab. de Physique Corpusculaire, IN2P3-CNRS,
     FR-75231 Paris Cedex 05, France
    \label{CDF}}
\titlefoot{CERN, CH-1211 Geneva 23, Switzerland
    \label{CERN}}
\titlefoot{Institut de Recherches Subatomiques, IN2P3 - CNRS/ULP - BP20,
     FR-67037 Strasbourg Cedex, France
    \label{CRN}}
\titlefoot{Now at DESY-Zeuthen, Platanenallee 6, D-15735 Zeuthen, Germany
    \label{DESY}}
\titlefoot{Institute of Nuclear Physics, N.C.S.R. Demokritos,
     P.O. Box 60228, GR-15310 Athens, Greece
    \label{DEMOKRITOS}}
\titlefoot{FZU, Inst. of Phys. of the C.A.S. High Energy Physics Division,
     Na Slovance 2, CZ-180 40, Praha 8, Czech Republic
    \label{FZU}}
\titlefoot{Dipartimento di Fisica, Universit\`a di Genova and INFN,
     Via Dodecaneso 33, IT-16146 Genova, Italy
    \label{GENOVA}}
\titlefoot{Institut des Sciences Nucl\'eaires, IN2P3-CNRS, Universit\'e
     de Grenoble 1, FR-38026 Grenoble Cedex, France
    \label{GRENOBLE}}
\titlefoot{Helsinki Institute of Physics, HIP,
     P.O. Box 9, FI-00014 Helsinki, Finland
    \label{HELSINKI}}
\titlefoot{Joint Institute for Nuclear Research, Dubna, Head Post
     Office, P.O. Box 79, RU-101 000 Moscow, Russian Federation
    \label{JINR}}
\titlefoot{Institut f\"ur Experimentelle Kernphysik,
     Universit\"at Karlsruhe, Postfach 6980, DE-76128 Karlsruhe,
     Germany
    \label{KARLSRUHE}}
\titlefoot{Institute of Nuclear Physics,Ul. Kawiory 26a,
     PL-30055 Krakow, Poland
    \label{KRAKOW1}}
\titlefoot{Faculty of Physics and Nuclear Techniques, University of Mining
     and Metallurgy, PL-30055 Krakow, Poland
    \label{KRAKOW2}}
\titlefoot{Universit\'e de Paris-Sud, Lab. de l'Acc\'el\'erateur
     Lin\'eaire, IN2P3-CNRS, B\^{a}t. 200, FR-91405 Orsay Cedex, France
    \label{LAL}}
\titlefoot{School of Physics and Chemistry, University of Lancaster,
     Lancaster LA1 4YB, UK
    \label{LANCASTER}}
\titlefoot{LIP, IST, FCUL - Av. Elias Garcia, 14-$1^{o}$,
     PT-1000 Lisboa Codex, Portugal
    \label{LIP}}
\titlefoot{Department of Physics, University of Liverpool, P.O.
     Box 147, Liverpool L69 3BX, UK
    \label{LIVERPOOL}}
\titlefoot{LPNHE, IN2P3-CNRS, Univ.~Paris VI et VII, Tour 33 (RdC),
     4 place Jussieu, FR-75252 Paris Cedex 05, France
    \label{LPNHE}}
\titlefoot{Department of Physics, University of Lund,
     S\"olvegatan 14, SE-223 63 Lund, Sweden
    \label{LUND}}
\titlefoot{Universit\'e Claude Bernard de Lyon, IPNL, IN2P3-CNRS,
     FR-69622 Villeurbanne Cedex, France
    \label{LYON}}
\titlefoot{Dipartimento di Fisica, Universit\`a di Milano and INFN-MILANO,
     Via Celoria 16, IT-20133 Milan, Italy
    \label{MILANO}}
\titlefoot{Dipartimento di Fisica, Univ. di Milano-Bicocca and
     INFN-MILANO, Piazza della Scienza 2, IT-20126 Milan, Italy
    \label{MILANO2}}
\titlefoot{IPNP of MFF, Charles Univ., Areal MFF,
     V Holesovickach 2, CZ-180 00, Praha 8, Czech Republic
    \label{NC}}
\titlefoot{NIKHEF, Postbus 41882, NL-1009 DB
     Amsterdam, The Netherlands
    \label{NIKHEF}}
\titlefoot{National Technical University, Physics Department,
     Zografou Campus, GR-15773 Athens, Greece
    \label{NTU-ATHENS}}
\titlefoot{Physics Department, University of Oslo, Blindern,
     NO-0316 Oslo, Norway
    \label{OSLO}}
\titlefoot{Dpto. Fisica, Univ. Oviedo, Avda. Calvo Sotelo
     s/n, ES-33007 Oviedo, Spain
    \label{OVIEDO}}
\titlefoot{Department of Physics, University of Oxford,
     Keble Road, Oxford OX1 3RH, UK
    \label{OXFORD}}
\titlefoot{Dipartimento di Fisica, Universit\`a di Padova and
     INFN, Via Marzolo 8, IT-35131 Padua, Italy
    \label{PADOVA}}
\titlefoot{Rutherford Appleton Laboratory, Chilton, Didcot
     OX11 OQX, UK
    \label{RAL}}
\titlefoot{Dipartimento di Fisica, Universit\`a di Roma II and
     INFN, Tor Vergata, IT-00173 Rome, Italy
    \label{ROMA2}}
\titlefoot{Dipartimento di Fisica, Universit\`a di Roma III and
     INFN, Via della Vasca Navale 84, IT-00146 Rome, Italy
    \label{ROMA3}}
\titlefoot{DAPNIA/Service de Physique des Particules,
     CEA-Saclay, FR-91191 Gif-sur-Yvette Cedex, France
    \label{SACLAY}}
\titlefoot{Instituto de Fisica de Cantabria (CSIC-UC), Avda.
     los Castros s/n, ES-39006 Santander, Spain
    \label{SANTANDER}}
\titlefoot{Inst. for High Energy Physics, Serpukov
     P.O. Box 35, Protvino, (Moscow Region), Russian Federation
    \label{SERPUKHOV}}
\titlefoot{J. Stefan Institute, Jamova 39, SI-1000 Ljubljana, Slovenia
     and Laboratory for Astroparticle Physics,\\
     \indent~~Nova Gorica Polytechnic, Kostanjeviska 16a, SI-5000 Nova Gorica, Slovenia, \\
     \indent~~and Department of Physics, University of Ljubljana,
     SI-1000 Ljubljana, Slovenia
    \label{SLOVENIJA}}
\titlefoot{Fysikum, Stockholm University,
     Box 6730, SE-113 85 Stockholm, Sweden
    \label{STOCKHOLM}}
\titlefoot{Dipartimento di Fisica Sperimentale, Universit\`a di
     Torino and INFN, Via P. Giuria 1, IT-10125 Turin, Italy
    \label{TORINO}}
\titlefoot{INFN,Sezione di Torino, and Dipartimento di Fisica Teorica,
     Universit\`a di Torino, Via P. Giuria 1,\\
     \indent~~IT-10125 Turin, Italy
    \label{TORINOTH}}
\titlefoot{Dipartimento di Fisica, Universit\`a di Trieste and
     INFN, Via A. Valerio 2, IT-34127 Trieste, Italy \\
     \indent~~and Istituto di Fisica, Universit\`a di Udine,
     IT-33100 Udine, Italy
    \label{TU}}
\titlefoot{Univ. Federal do Rio de Janeiro, C.P. 68528
     Cidade Univ., Ilha do Fund\~ao
     BR-21945-970 Rio de Janeiro, Brazil
    \label{UFRJ}}
\titlefoot{Department of Radiation Sciences, University of
     Uppsala, P.O. Box 535, SE-751 21 Uppsala, Sweden
    \label{UPPSALA}}
\titlefoot{IFIC, Valencia-CSIC, and D.F.A.M.N., U. de Valencia,
     Avda. Dr. Moliner 50, ES-46100 Burjassot (Valencia), Spain
    \label{VALENCIA}}
\titlefoot{Institut f\"ur Hochenergiephysik, \"Osterr. Akad.
     d. Wissensch., Nikolsdorfergasse 18, AT-1050 Vienna, Austria
    \label{VIENNA}}
\titlefoot{Inst. Nuclear Studies and University of Warsaw, Ul.
     Hoza 69, PL-00681 Warsaw, Poland
    \label{WARSZAWA}}
\titlefoot{Fachbereich Physik, University of Wuppertal, Postfach
     100 127, DE-42097 Wuppertal, Germany \\
\noindent
{$^\dagger$~deceased}
    \label{WUPPERTAL}}

\addtolength{\textheight}{-10mm}
\addtolength{\footskip}{5mm}
\clearpage
\headsep 30.0pt
\end{titlepage}
%
\pagenumbering{arabic} 
\setcounter{footnote}{0} %
\large

\section{Introduction}

Decays~$b\rightarrow\bar c$ are expected to occur through the
Cabibbo favoured transitions~$b\rightarrow cW^-$ and
$W^-\rightarrow\bar cs$~\footnote{Charge conjugate reactions are
  implied throughout this paper.}. 
Hence, neglecting $b\rightarrow u$ transition and $D^0$ mixing, $b$-hadron decays to wrong
sign charmed mesons are in fact double charm transitions. The double
charm rate is related to $n_c$,
the mean number of charm quarks (and anti-quarks) produced per
$b$-decay:
\begin{equation}
  n_c=1-{\mathcal B}(b\rightarrow{\mathrm no~open~charm})+2{\mathcal
    B}(b\rightarrow{\mathrm charmonium})+{\mathcal B}(b\rightarrow
    {\mathrm double~charm})
\end{equation}
which can be predicted by Heavy Quark Effective Theory (HQET) based
calculations of the semileptonic $B$~meson branching
fraction~\cite{ref:0}.

Evidence for wrong sign charm production and double charmed $b$-decays
has been found both at the $\Upsilon(4S)$ and at
LEP. ARGUS~\cite{ref:1} and CLEO~\cite{ref:2} have shown
evidence for the two-body transitions~$B\rightarrow D_s^{(*)+}\bar
D^{(*)}$~\footnote{In the following, $D$ ($D_{(s)}$) denotes either
  $D^0$ or $D^+$ ($D^0$, $D^+$ or $D_s^+$).}. From the analysis of the
$D_s$~momentum spectrum, these decays are found to contribute about
half of the total $D_s$~production at the $\Upsilon(4S)$, the
remainder coming from either $B\rightarrow D_s^{(*)+}\bar D^{**}$ or
$B\rightarrow D_s^{(*)+}\bar D^{(*)}\pi,\rho,\omega$ (where $D^{**}$
denotes an orbitally excited $D$~meson). By using $D$-lepton
correlations, CLEO has observed wrong sign~$D$
production~\cite{ref:2b}. ALEPH has reported evidence for
$b\rightarrow D\bar D_{(s)}X$~decays with both charmed mesons
reconstructed~\cite{ref:3}. The observed $D\bar DX$~signal is shown to
originate either from $B\rightarrow D^{(*)}\bar D^{(*)}K^{(*)}$ or
from $B\rightarrow D_s^{**}\bar D$ with a subsequent decay of the
orbitally excited state~$D_s^{**}$ into $D^{(*)}K$.

In this paper, the DELPHI data are used to measure the inclusive
branching fractions of $b$-hadrons into wrong sign charm mesons,
${\mathcal B}(b\rightarrow\bar DX)$ and ${\mathcal
  B}(b\rightarrow D_s^-X)$. $D^0$, $D^+$ and $D_s^+$~mesons are
exclusively reconstructed in $Z\rightarrow b\bar b$~events, recorded
by DELPHI in the years~1994 and 1995. The wrong sign contribution is
extracted by using two discriminant variables: the charge of the
$b$-quark at decay time, estimated from the charges of identified
particles, and the momentum of the charmed meson in the rest frame of
the $b$-hadron.

\section{Experimental procedure}

\subsection{The DELPHI detector}

A detailed description of the DELPHI detector and its performance can
be found in reference~\cite{ref:4}. Only the subdetectors relevant to
the present analysis are described in the following.

The tracking of charged particles in the barrel region is accomplished with
a set of cylindrical tracking detectors whose axes are oriented along the
1.23~T magnetic field and the direction of the beam.

The Time Projection Chamber (TPC), the main tracking device, is a cylinder
of 30~cm inner radius, 122~cm outer radius and a length of 2.7~m. For polar
angles between 39$^\circ$ and 141$^\circ$, it provides up to 16~space
points along the charged particle trajectory~\footnote{In the DELPHI
  frame, the $z$~axis is defined along the electron beam direction,
  the $x$~axis points towards the centre of
  the LEP ring and the $y$~axis points upwards. The polar angle to
  the $z$~axis is called~$\theta$; the azimuthal angle around the
  $z$~axis is referred to as~$\phi$. The radial coordinate is
  $R=\sqrt{x^2+y^2}$.}.

The Vertex Detector (VD), located nearest to the LEP interaction region,
consists of three concentric layers of silicon microstrip detectors at
average radii of 6.3~cm, 9.0~cm and 10.9~cm. Since 1994, the innermost
and the outermost layers were equipped with double sided silicon
microstrip modules allowing both $R\phi$ and $z$~measurements.

Hadrons are identified using the specific ionization
(${\mathrm d}E/{\mathrm d}x$) measured in the TPC and the Cherenkov
radiation detected in the barrel Ring Imaging CHerenkov counter (RICH)
placed between the TPC and the Outer Detector (OD).

\subsection{Event sample}

For this analysis, the data collected by the DELPHI~experiment
in the years 1994 and 1995 at $\sqrt s$ close to 91.2~GeV are used, 
corresponding to about 2.1~million hadronic $Z$~decays. Simulated hadronic
events are generated with the JETSET~7.3 program~\cite{ref:5}. Full
detector simulation is applied to Monte Carlo events which are
subsequently processed through the same analysis chain as the real
data~\cite{ref:4}.

The decays~$B\rightarrow D_s^{(*)+}\bar D^{(*)}$, $B\rightarrow
D_s^{(*)+}\bar D^{**}$, $B\rightarrow D_s^{(*)+}\bar
D^{(*)}\pi,\rho,\omega$, $B\rightarrow D_s^{**}\bar D\rightarrow
D^{(*)}K\bar D$ and $B\rightarrow D^{(*)}\bar D^{(*)}K^{(*)}$ are used
to model $b$-decay into wrong sign charmed mesons. The $b$-hadron decay
properties to right sign charm are adjusted to match the latest
experimental values~\cite{ref:10}. In total, a
sample of about 58,000~$b\rightarrow\bar D_{(s)}X$ and about
99,000~$b\rightarrow D_{(s)}X$~events, with $D_{(s)}$ forced to decay
into the modes used in the analysis, has been generated. The
background is modelled with about 3.2~million~$Z\rightarrow q\bar
q$ and about 1.8~million~$Z\rightarrow b\bar b$ Monte Carlo events.

Hadronic $Z$~decays are selected by requiring at least five charged
particles and a total charged energy larger than 12\% of the collision
energy~\cite{ref:4}. The tagging of $b$-quark jets is based on four
discriminant variables, the most important one being the probability
for all tracks to originate from the primary interaction vertex,
calculated from the track impact parameters with respect to this
point~\cite{ref:6}. The other variables are defined for jets with a
secondary vertex: effective mass of the system of particles attached
to the secondary vertex, rapidity of these particles with respect to the
jet direction and fraction of the charged energy of the jet included
in the secondary vertex. 
All jet $b$-tags in the event are combined and the cut on the event
probability is chosen such that about 90\% of the reconstructed
charmed mesons originate from $b$-hadron decay. Correspondingly, the
$Z\rightarrow b\bar b$~selection efficiency varies between 58\% and
74\% for the different charm modes.

Each selected event is divided into two hemispheres by
the plane orthogonal to the axis of the most energetic jet and passing
through the primary interaction point.

\subsection{Charmed meson reconstruction} \label{sec:2_3}

Charged particles are selected as follows: momentum larger than
100~MeV/$c$, relative error on the momentum measurement smaller than
100\% and $R\phi$ ($z$) impact parameter with respect to the primary
interaction vertex smaller than 4~cm (4~cm/$\sin\theta$).

Charmed mesons are searched for in the decay modes 
$D^0\rightarrow K^-\pi^+$, $D^+\rightarrow K^-\pi^+\pi^+$ and
$D_s^+\rightarrow\phi\pi^+\rightarrow K^+K^-\pi^+$ by trying all
possible combinations of charged particles in the hemisphere. The
${\mathrm d}E/{\mathrm d}x$~values of the kaon and pion candidates are
required to be consistent with the respective mass hypotheses. For
$D^+\rightarrow K^-\pi^+\pi^+$~decays which suffer from a high level of
combinatorial background, the kaon must be tagged additionally by the
RICH. To allow for a precise reconstruction of the $D_{(s)}$~decay
vertex, at least two tracks in each combination are required to have
associated hits in the Vertex Detector.

Track combinations satisfying these criteria are fitted to a common
vertex. The $\chi^2$-probability of the fit must exceed
0.01\%. Combinations containing a fragmentation track are rejected by
requiring the $D$ ($D_s$) vertex to lie at least three (two) standard
deviations away from the primary interaction point and imposing the
requirement~$x_E>0.15$ on the energy
fraction~$x_E=E_{D_{(s)}}/E_{beam}$. For
$D_s^+\rightarrow\phi\pi^+$, a selection at $\pm 12$~MeV/$c^2$ around the
nominal $\phi$~mass is applied to the reconstructed $K^+K^-$~mass.

$D$~candidates are selected by using four discriminant variables: the
RICH~information for the kaon candidate, the decay length from the
primary to the charm vertex divided by its error, the energy
fraction~$x_E$ and the cosine of the charm decay
angle~$\theta_D$, defined as the angle between
the $K$~momentum vector in the $D$ meson rest frame and the $D$~momentum
vector in the laboratory frame. The
$\cos\theta_D$~distribution is flat for the signal and peaked
at $-1$ for the combinatorial background. For $D_s$~candidates, two
additional variables are used: the reconstructed $K^+K^-$~mass and the
cosine of the $\phi$~helicity
angle~$\theta_H$. The latter is defined as the angle
between the $K^+$ and the $D_s$~direction in the $\phi$~rest
frame. The signal follows a $\cos^2\theta_H$~distribution
while the background is flat in $\cos\theta_H$. The different
variables~$x_i$ are combined by using a likelihood ratio:
\begin{equation}
  X(D_{(s)})=\frac{R(D_{(s)})}{1+R(D_{(s)})}~, \quad
  R(D_{(s)})=\prod_i\frac{S_i(x_i;D_{(s)})}{B_i(x_i;D_{(s)})}
\end{equation}
where $S_i$ and $B_i$ are the normalised distributions of $x_i$ for the signal
and the combinatorial background, respectively, 
as obtained from the simulation.
The combined variable is defined such that the target value is $X=1$ for
the signal and $X=0$ for the background. 
For each decay mode, the selection cut on the
variable~$X$ is adjusted on simulated events to optimise the
statistical significance of the signal. The following selections are found:
$X(D^0)>0.8$, $X(D^+)>0.6$ and $X(D_s)>0.95$.

\begin{figure}[p]
\begin{center}
\begin{sideways}
\begin{minipage}{\textheight}
  \begin{center}
    \includegraphics{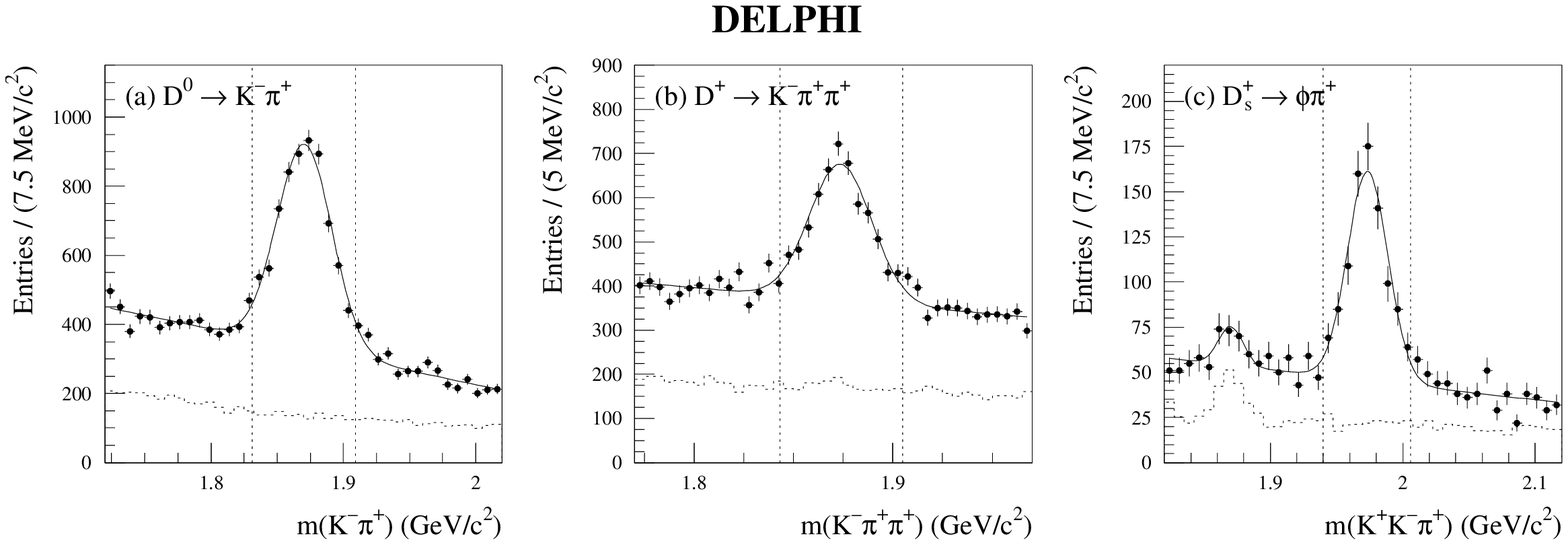}
    \caption{The invariant mass of selected $D^0\rightarrow K^-\pi^+$,
      $D^+\rightarrow K^-\pi^+\pi^+$ and
      $D_s^+\rightarrow\phi\pi^+\rightarrow
      K^+K^-\pi^+$~candidates. The points with error bars are the real
      data. The solid line is the result of the fit mentioned in the
      text. The signal window is shown by dashed vertical lines and the
      dashed histogram represents the Monte Carlo expectation for
      the combinatorial background (arbitrary normalisation).}
      \label{fig:1}
  \end{center}
\end{minipage}
\end{sideways}
\end{center}
\end{figure}

For each selected candidate, the invariant mass of the track
combination is computed (Figure~\ref{fig:1}). 7345 (6906, 984) $D^0$
($D^+$, $D_s$) candidates
are found in the signal window
corresponding to an interval of about $\pm 2\sigma$ around the signal
peak. The remaining combinatorial
background is determined by a fit to the invariant mass distribution. 
The fit uses a Gaussian function for the signal
and a linear parametrisation for the combinatorial background. The
satellite peak, due to
the $D^+\rightarrow K^+K^-\pi^+$ decay, in the
$D_s^+\rightarrow\phi\pi^+$~channel is also fitted by a Gaussian
function. In this way, the combinatorial background is found to be
$3038\pm 43$, $4677\pm 66$ and $404\pm 12$ for $D^0\rightarrow
K^-\pi^+$, $D^+\rightarrow K^-\pi^+\pi^+$ and
$D_s^+\rightarrow\phi\pi^+$, respectively.

\subsection{Discriminant variables}

The discriminant variables used for selecting wrong sign decays are
constructed by using a common DELPHI analysis package called
BSAURUS. Details on how the different BSAURUS variables are formed can
be found in reference~\cite{ref:7}.

The flavour of the $D_{(s)}$~meson, \textit{i.e.}, the charge of the
$c$-quark confined in the charmed meson, is determined from the charge
of the kaon for the channels $D^0\rightarrow K^-\pi^+$ and
$D^+\rightarrow K^-\pi^+\pi^+$, and from the charge of the pion for
the channel $D_s^+\rightarrow\phi\pi^+$.

The charge of the $b$-quark at decay time
in the hemisphere of the charmed meson is obtained from the BSAURUS
Decay Flavour Neural Network (BDFNN). The approach is to first form the
weighted sum of particle charges in the hemisphere excluding the particles
from the exclusive decay of the $D_{(s)}$, in order to avoid a possible
bias. The weighting factor is constructed from the conditional
probability for the track to have the same charge as the decaying
$b$-quark, and is determined via a neural network technique based
mainly on particle identification variables for kaons, protons,
electrons and muons and $B-D$~vertex separation variables. In order to
use optimally the event information, the resulting hemisphere charges
are constructed separately to estimate the $b$-charge at both
production and decay time and this is repeated for each of the
$b$-hadron types ($B^+,B^0,B_s^0, b$-baryon). In a final step, these
hemisphere charges form the main input variables to a neural network
trained to find the $b$-quark charge in combination with BSAURUS
$b$-hadron type tagging probabilities and also including charge
correlation information from the opposite hemisphere.

The wrong sign tag~$Y(D_{(s)})$, the first discriminant variable,
is obtained by correlating the BDFNN output with the
flavour of the charmed meson:
\begin{equation}
  Y(D_{(s)})=\left\{\begin{array}{ll}
  {\mathrm BDFNN} & \textrm{for $D_{(s)}$}\\
  1-{\mathrm BDFNN} & \textrm{for $\bar D_{(s)}$}
  \end{array}\right.~.
\end{equation}
The target values of the BSAURUS Decay Flavour Neural Network are
${\mathrm BDFNN}=1$ and ${\mathrm BDFNN}=0$ for $b$ and $\bar
b$-hadrons, respectively. Hence, the target value of the wrong sign
tag is $Y=0$ for wrong sign decays and $Y=1$ for right sign.

While wrong sign and $B_s$~right sign decays contribute about equally
to the $D_s$~production in $Z\rightarrow b\bar b$~events, the wrong
sign production mechanism is strongly suppressed in the case of
$D$~mesons. Hence, for selecting wrong sign $D^0$ and $D^+$~mesons,
stronger discrimination is required and a second variable, the
momentum of the $D$~meson in the $b$-hadron rest frame, $p(D)$, is
used.

The $b$-hadron four-momentum in the hemisphere of the charmed meson
is inclusively reconstructed in BSAURUS using the following
procedure. An initial estimate of the $b$-hadron momentum~$\vec
p_{raw}$, energy~$E_{raw}$ and mass
$m_{raw}=\sqrt{E_{raw}^2-p_{raw}^2}$ is formed from
particles with high rapidity for events with more than two-jets and from
the sum of ``$b$-weighted'' four-vectors for the two-jet case. This
weighting involves the use of neural networks trained to identify
tracks originating from the weakly decaying $b$-hadron in the
hemisphere. $E_{raw}$~is then corrected, hemisphere-by-hemisphere,
motivated by the observation (in Monte Carlo simulation) of a
correlation between the energy residuals $\Delta E=E_{raw}-E_{true}$
and $m_{raw}$ (which is approximately linear in $m_{raw}$)
and a further correlation between $\Delta E$ and
$x_h=E_{hem}/E_{beam}$, where $E_{hem}$ is the sum of the energies of
all particles reconstructed in the hemisphere, resulting from neutral
energy losses and inefficiencies. These effects are parametrised and
corrected for, after which the resolution obtained in $p(D)$ is about
$\pm$300~MeV/$c$.

\begin{figure}[p]
  \begin{center}
    \includegraphics{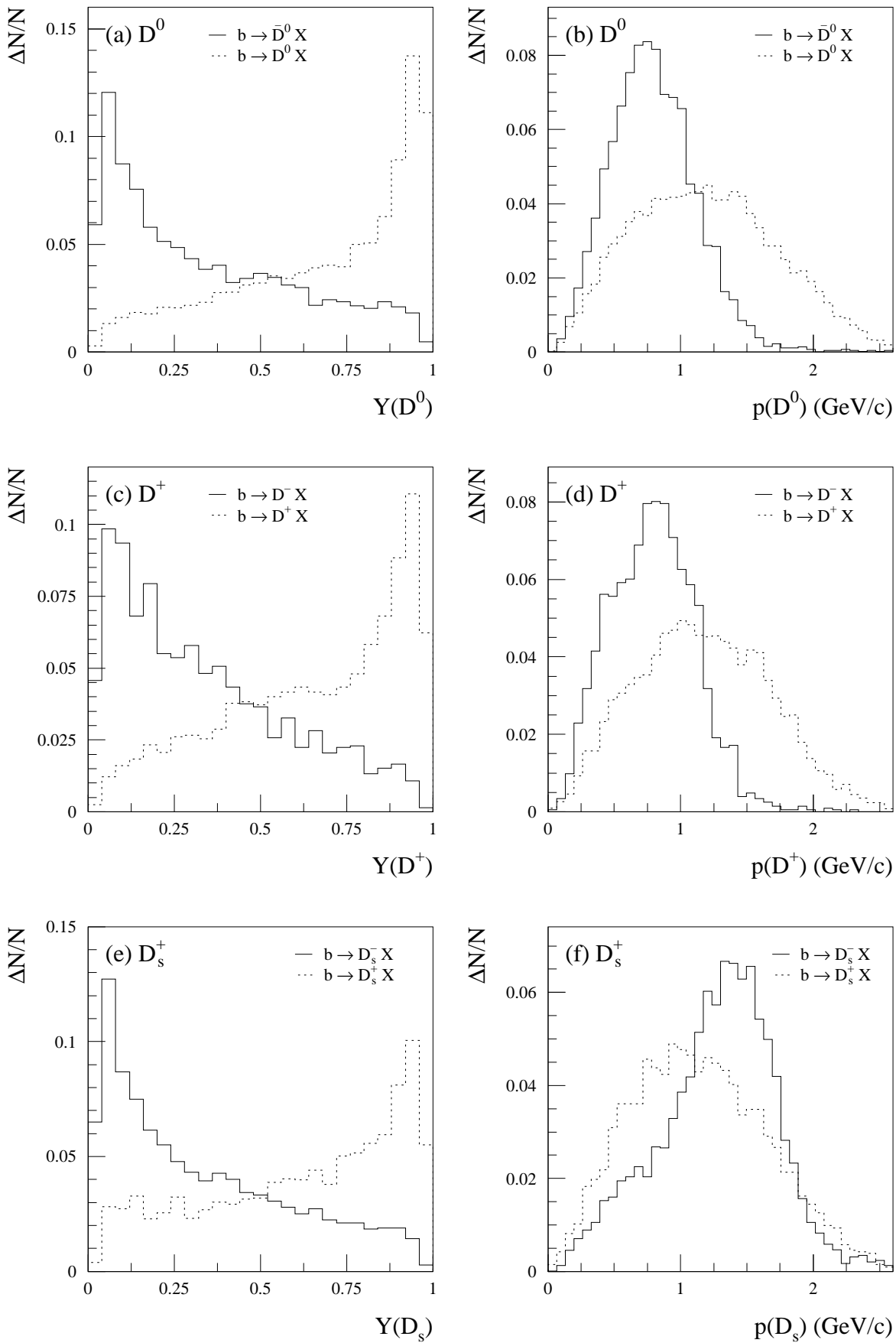}
    \caption{The wrong sign tag~$Y(D_{(s)})$ and the momentum of the
      charmed meson in the rest frame of the $b$-hadron~$p(D_{(s)})$,
      shown for wrong sign and right sign Monte Carlo events. The
      number of entries in each histogram is normalised to one.}
    \label{fig:2}
  \end{center}
\end{figure}

The two discriminant variables are shown in Figure~\ref{fig:2}.

\subsection{The fit}

For each charm decay mode, the numbers of wrong sign and right
sign events, $N_W$ and $N_R$, are
determined by a fit to the above-mentioned discriminant variables. The
following components can contribute to the distributions of these
variables in the real data: wrong sign $b\rightarrow\bar
D_{(s)}X$~mesons, right
sign $b\rightarrow D_{(s)}X$~mesons, $D_{(s)}$~meson background
(contamination by charmed mesons produced in $Z\rightarrow c\bar
c$~events) and combinatorial background. The shapes of
the distributions of these four components ($F^W$, $F^R$, $F^{c\bar
  c}$ and $F^{Bkgrd}$) are determined
from the Monte Carlo simulation. In each fit, the number of charmed
mesons from $Z\rightarrow c\bar c$~events is fixed to the value calculated
from the partial
width~$R_c=0.1702\pm 0.0034$, the fragmentation
probabilities~$f(c\rightarrow D^0)=0.552\pm 0.037$, $f(c\rightarrow
D^+)=0.237\pm 0.016$, $f(c\rightarrow D_s)=0.121\pm
0.025$~\cite{ref:8} and the acceptance determined from the
simulation. The numbers are found to be $436\pm 30$, $266\pm 19$ and
$73\pm 15$ for $D^0\rightarrow K^-\pi^+$, $D^+\rightarrow K^-\pi^+\pi^+$ and
$D_s^+\rightarrow\phi\pi^+$, respectively. The normalisation of the
combinatorial background is fixed to the values quoted in
Section~\ref{sec:2_3}.

Selected $D_s$~meson candidates are arranged in 10~bins,
$i$, of $Y(D_s)$ (bin width~0.1) and the resulting one-dimensional
histogram is fitted by the function:
\begin{equation}
  N_i=N_W F_i^W+N_R F_i^R+N_{c\bar c}F_i^{c\bar c}+N_{Bkgrd}F_i^{Bkgrd}~.
\end{equation}
The normalisations $\Sigma_i F_i^W=1$, $\Sigma_i F_i^R=1$, $\Sigma_i
F_i^{c\bar c}=1$ and $\Sigma_i F_i^{Bkgrd}=1$ are used. Selected $D^0$
and $D^+$~candidates are arranged in 4~bins, $i$, of $Y(D)$ (bin
width~0.25) and 13~bins, $j$, of $p(D)$ (bin width~200~MeV/$c$) and the
fit function:
\begin{equation}
  N_{ij}=N_W F_{ij}^W+N_R F_{ij}^R+N_{c\bar c}F_{ij}^{c\bar
  c}+N_{Bkgrd}F_{ij}^{Bkgrd}
\end{equation}
is used. The fit algorithm accounts for finite Monte Carlo
statistics~\cite{ref:9} and the total number of selected candidates is
used as a constraint. By applying the algorithm to simulated
$Z\rightarrow q\bar q$ events, no significant bias in the fit result
is observed.

\begin{figure}[p]
  \begin{center}
    \includegraphics{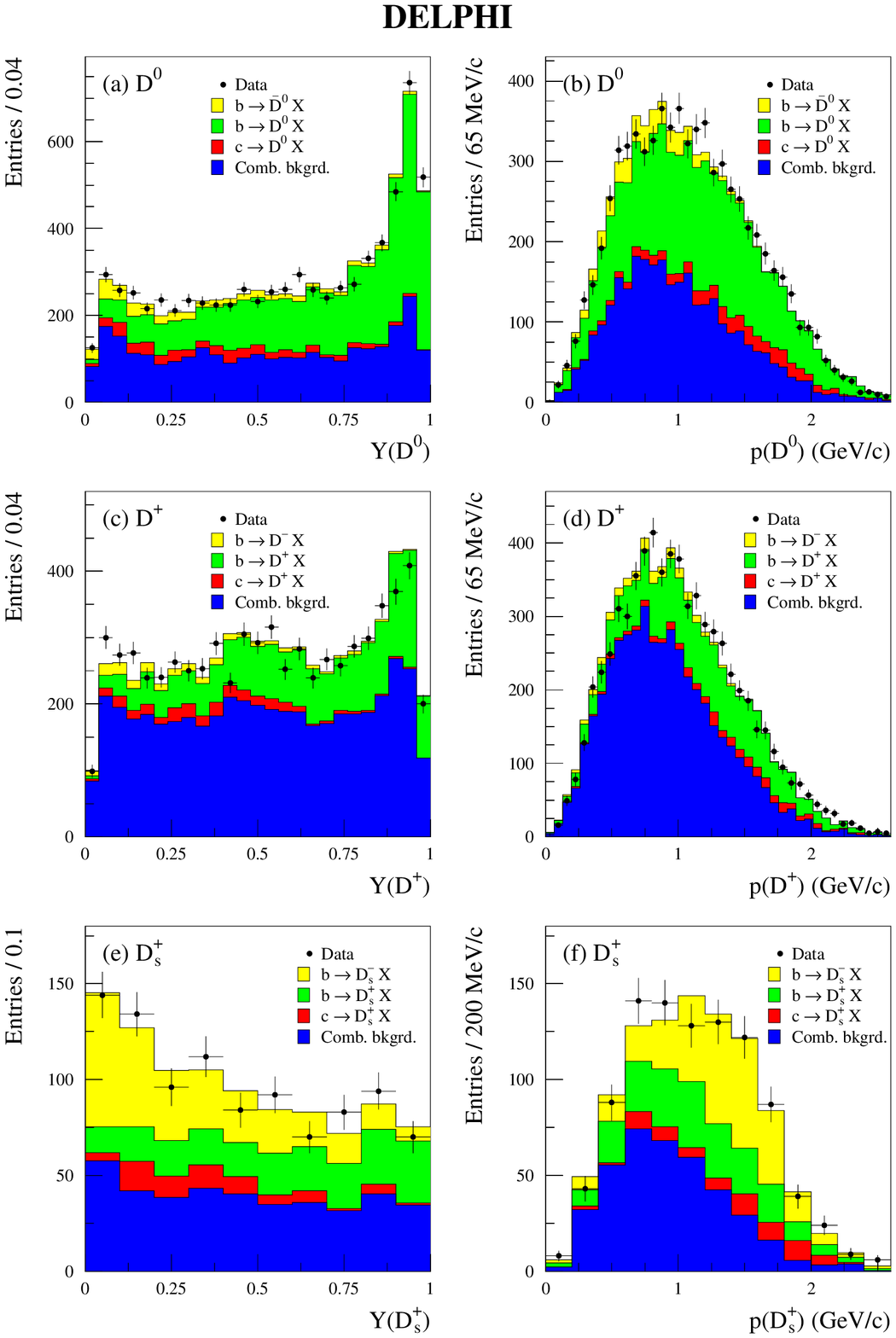}
    \caption{The wrong sign tag~$Y(D_{(s)})$ and the momentum of the
      charmed meson in the rest frame of the
      $b$-hadron~$p(D_{(s)})$. The data are the points with error
      bars; the histograms are the components of the fit function
      (as described in the text).} \label{fig:3}
  \end{center}
\end{figure}

\begin{figure}[p]
  \begin{center}
    \includegraphics{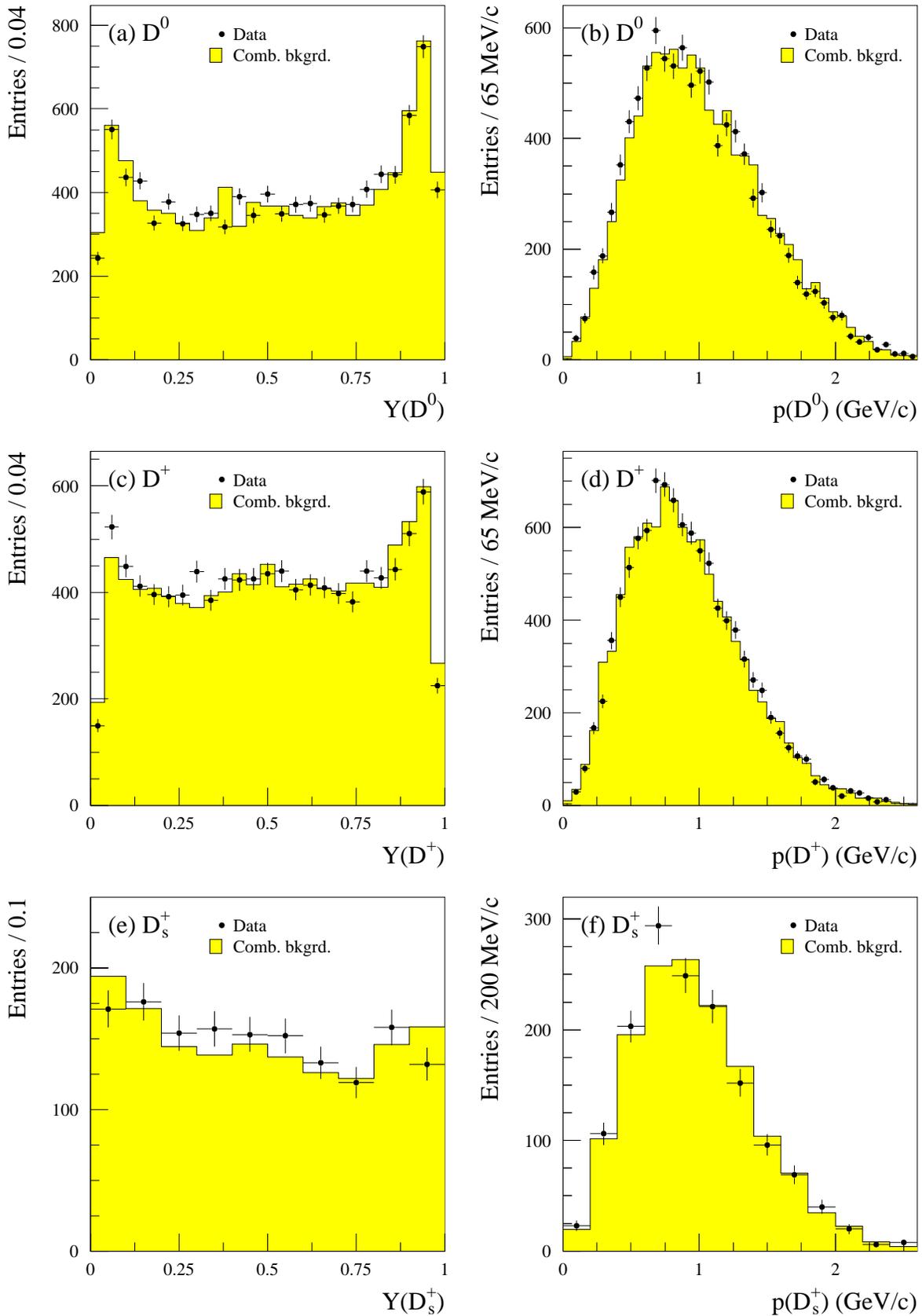}
    \caption{Same as Figure~\ref{fig:3} for candidates selected
      outside the signal mass window.} \label{fig:4}
  \end{center}
\end{figure}

\begin{figure}[p]
  \begin{center}
    \includegraphics{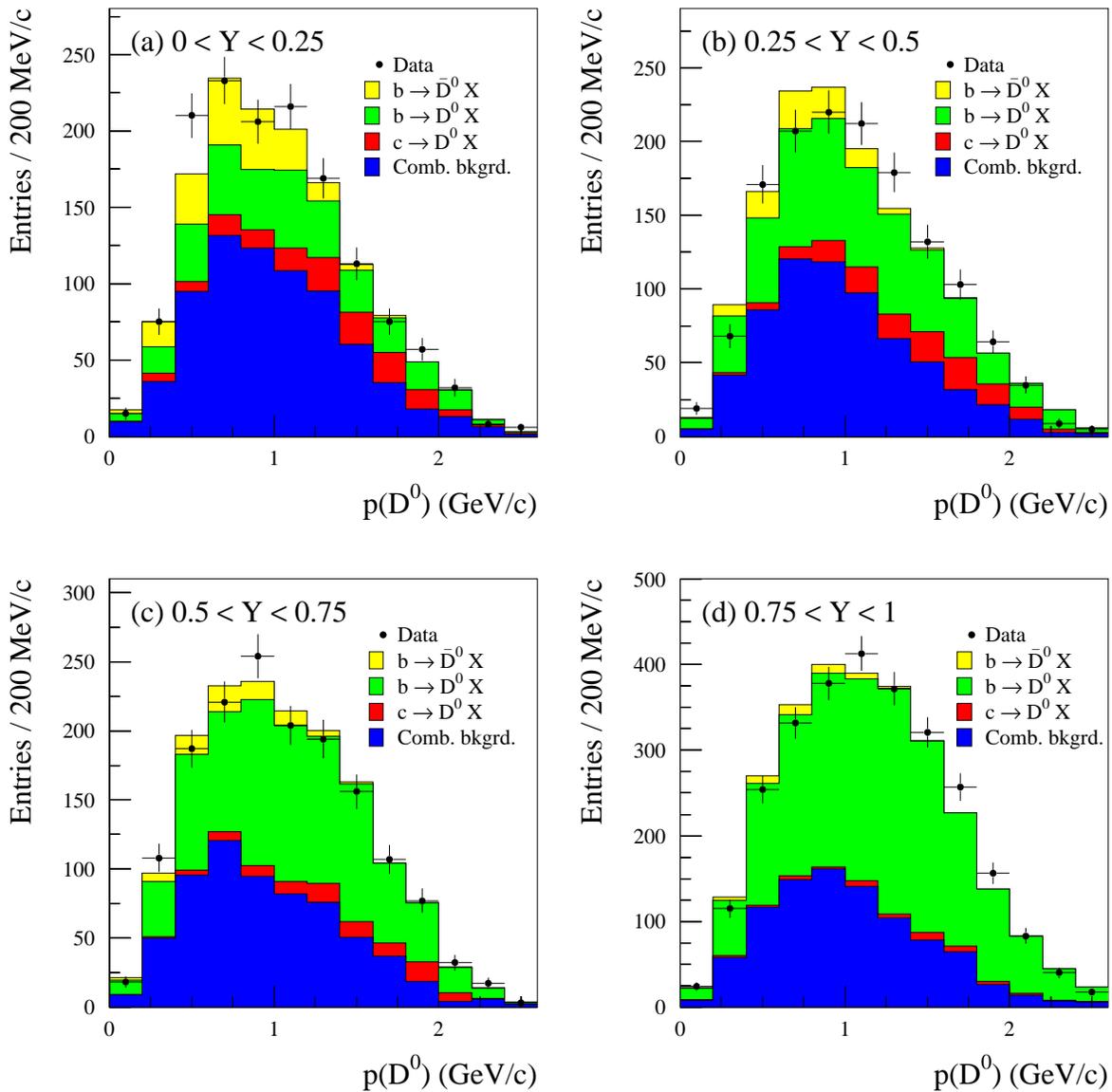}
    \caption{The $D^0$~momentum in the $b$-hadron rest frame $p(D^0)$
      in bins of the wrong sign tag~$Y(D^0)$. The data are the points
      with error bars; the histograms are the components of the
      fit function (as described in the text).} \label{fig:5}
  \end{center}
\end{figure}

\begin{figure}[p]
  \begin{center}
    \includegraphics{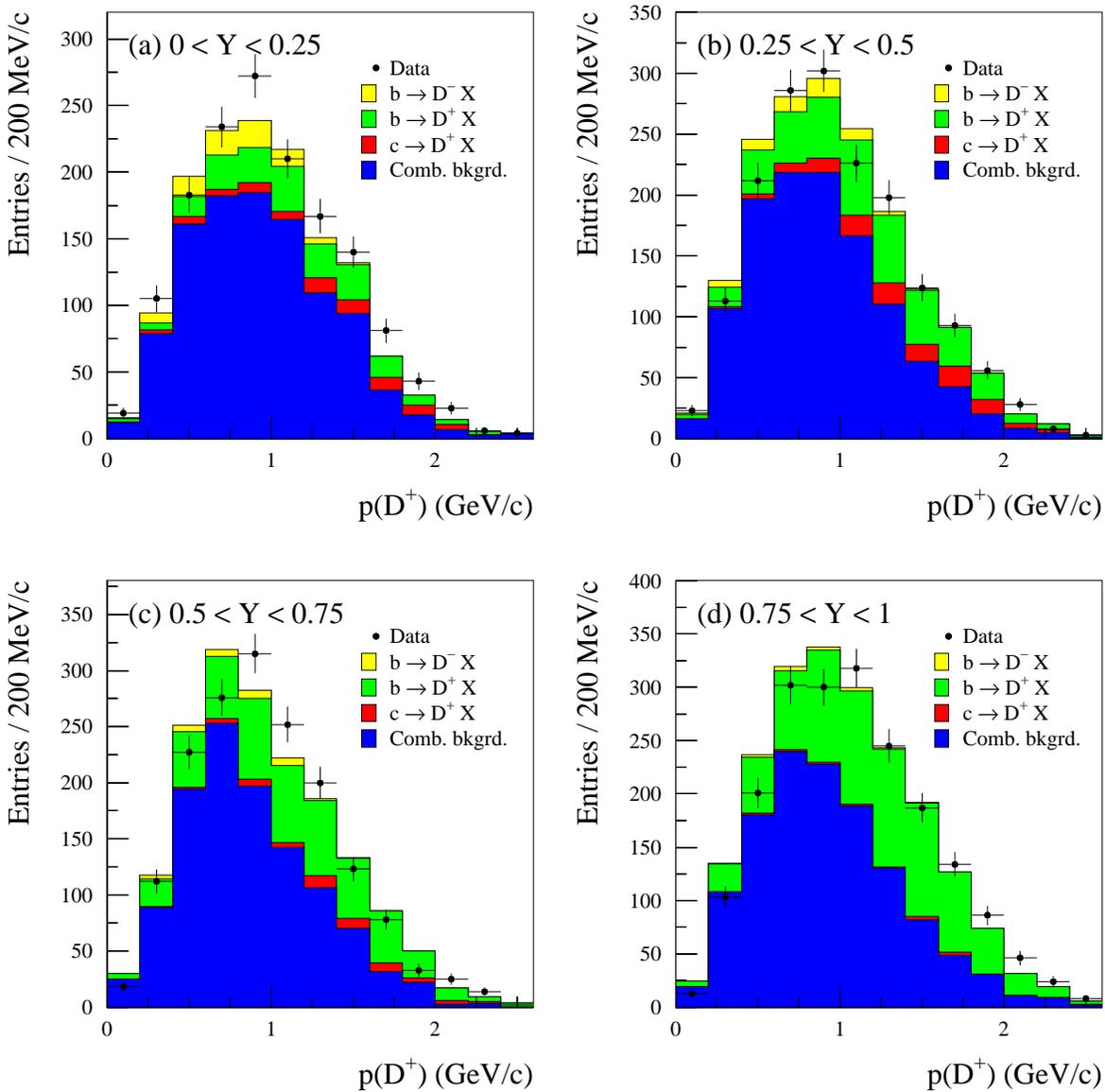}
    \caption{Same as Figure~\ref{fig:5} for~$D^+$.} \label{fig:6}
  \end{center}
\end{figure}

The results obtained by fitting the real data are shown in
Figures~\ref{fig:3}, \ref{fig:5} and \ref{fig:6}. The Monte Carlo
model of the combinatorial background is tested on real data
$D_{(s)}$~candidates selected outside the signal mass window
(Figure~\ref{fig:4}). The numbers of wrong sign and right sign events
for each decay channel are given in Table~\ref{tab:1}. For the
one-dimensional fit, the value of the $\chi^2$ is 4.8 compared to
$10-1$~degrees of freedom. The two-dimensional fit has $52-1$ degrees
of freedom and the $\chi^2$ is 52.0 (62.3) for the $D^0$ ($D^+$) sample.
\begin{table}
  \begin{center}
    \begin{tabular}{|c|c|c|c|c|}
      \hline
      \rule[-2.3ex]{0pt}{6ex}Sample & Wrong sign evts. &
      Right sign evts. & $\epsilon_W/\epsilon_R$ &
      $\frac{{\mathcal B}(b\rightarrow\bar
      D_{(s)}X)}{{\mathcal B}(b\rightarrow D_{(s)},\bar D_{(s)}X)}$ (\%)\\
      \hline \hline
      \rule{0pt}{2.7ex}$D^0\rightarrow K^-\pi^+$ & $383\pm 81$ &
      $3,396\pm 110$ & $0.92\pm 0.02$ & $11.0\pm 2.1\pm 1.5$\\
      $D^+\rightarrow K^-\pi^+\pi^+$ & $186\pm 86$ & $1,811\pm 101$ &
      $0.80\pm 0.03$ & $11.4\pm 4.7\pm 3.5$\\
      \rule[-1.3ex]{0pt}{1.3ex}$D_s^+\rightarrow \phi\pi^+$ & $286\pm
      42$ & $221\pm 39$ & $1.01\pm 0.03$ & $56.2\pm 5.7\pm 3.3$\\
      \hline
    \end{tabular}
    \caption{The fitted numbers of wrong sign and right sign mesons,
      the relative selection efficiency of wrong sign and
      right sign mesons and the fraction of wrong sign events in the
      charm signal. The error on the number of events is purely
      statistical. The error quoted on
      $\epsilon_W/\epsilon_R$ is just that due to
      Monte Carlo statistics. The first error on the wrong sign
      fraction is statistical; the second one is the sum of all
      systematic uncertainties listed in Table~\ref{tab:2}.}
      \label{tab:1}
  \end{center}
\end{table}

\section{Results and systematic uncertainties}

For each charm decay mode, the fraction of wrong sign
events~$b\rightarrow\bar D_{(s)}X$ in the signal~$b\rightarrow
D_{(s)},\bar D_{(s)}X$ is calculated from the numbers of wrong sign
and right sign events:
\begin{equation} \label{eq:1}
  \frac{{\mathcal B}(b\rightarrow\bar D_{(s)}X)}{{\mathcal
  B}(b\rightarrow D_{(s)},\bar D_{(s)}X)}=\frac{N_W}
  {N_W+(\epsilon_W/\epsilon_R)N_R}~.
\end{equation}
The results are given in Table~\ref{tab:1}. The
factor~$\epsilon_W/\epsilon_R$ in the denominator of
Eq.~\ref{eq:1} corrects for the different selection efficiencies of
wrong sign and right sign mesons and was obtained from the simulation. 

The wrong sign charm model used for the fit assumes a 50\%~contribution of
two-body decays~$B\rightarrow D_s^{(*)+}\bar D^{(*)}$ to the wrong
sign $D_s$~signal and the same relative contribution of $B\rightarrow
D_s^{**}\bar D$ to wrong sign~$D$. The corresponding modelling
uncertainty (Table~\ref{tab:2}) is estimated by varying these ratios
within $(50\pm 13)\%$ and $(50\pm 25)\%$, respectively.
\begin{table}[p]
\begin{center}
\begin{sideways}
\begin{minipage}{\textheight}
  \begin{center}
    \begin{tabular}{|l|c|c|c|c|c|}
      \hline
      \rule[-2.3ex]{0pt}{6ex}Source & Value &
      $\Delta\frac{{\mathcal B}(b\rightarrow\bar
      D^0X)}{{\mathcal B}(b\rightarrow D^0,\bar D^0X)}$ (\%) &
      $\Delta\frac{{\mathcal B}(b\rightarrow
      D^-X)}{{\mathcal B}(b\rightarrow D^\pm X)}$ (\%) &
      $\Delta\frac{{\mathcal B}(b\rightarrow
      D_s^-X)}{{\mathcal B}(b\rightarrow D_s^\pm X)}$ (\%) & Ref.\\
      \hline \hline
      \rule{0pt}{2.7ex}Model dependence (w.s.) & & & & & \\
      \quad $B\rightarrow D_s^{**}\bar D$~fraction & $(50\pm 25)\%$ & 0.12 &
      1.15 & & \cite{ref:3,ref:9b}\\
      \quad $B\rightarrow D_s^{(*)+}\bar D^{(*)}$~fraction & $(50\pm
      13)\%$ & & & 0.54 & \cite{ref:1,ref:2}\\
      Model dependence (r.s.) & & & & & \\
      \quad ${\mathcal B}(b\rightarrow D^0l^-\bar\nu X)$ & $(6.60\pm
      0.60)\%$ & 0.20 & & & \cite{ref:3b}\\
      \quad ${\mathcal B}(b\rightarrow D^+l^-\bar\nu X)$ & $(2.02\pm
      0.29)\%$ & & 0.61 & & \cite{ref:3b}\\
      \quad ${\mathcal B}(b\rightarrow D_s^+l^-\bar\nu X)$ & $(0.87\pm 0.28)\%$
      & & & 0.61 & \cite{ref:10}\\
      \quad ${\mathcal B}(b\rightarrow D^0D_s^-X)$ & $(9.10\pm 3.35)\%$ &
      0.08 & & & \cite{ref:3}\\
      \quad ${\mathcal B}(b\rightarrow D^+D_s^-X)$ & $(4.00\pm 2.05)\%$ &
      & 0.59 & & \cite{ref:3}\\
      \quad ${\mathcal B}(b\rightarrow D_s^+D_s^-X)$ & $(1.17\pm 0.71)\%$ &
      & & 2.09 & \cite{ref:3}\\
      \quad ${\mathcal B}(b\rightarrow D^0\bar DX)$ & $(6.45\pm
      2.08)\%$ & 1.40 & & & \cite{ref:3}\\
      \quad ${\mathcal B}(b\rightarrow D^+\bar DX)$ & $(1.80\pm
      0.96)\%$ & & 2.18 & & \cite{ref:3}\\
      \quad ${\mathcal B}(b\rightarrow D_s^+\bar DX)$ & $(1.17\pm
      0.71)\%$ & & & 2.39 & \\
      $Z\rightarrow c\bar c$~background & & 0.09 & 0.05 & 0.56 & \\
      Combinatorial background & & & & & \\
      \quad normalisation & & 0.37 & 1.46 & 0.29 & \\
      \quad shape & & & 1.56 & & \\
      \rule[-1.3ex]{0pt}{1.3ex}$\epsilon_W/\epsilon_R$ & & 0.41 & 1.04 &
      0.13 & \\
      \hline \hline
      \rule[-1.3ex]{0pt}{4ex}Total & & 1.53 & 3.53 & 3.34 & \\
      \hline
    \end{tabular}
    \caption{Breakdown of the systematic error on the wrong sign
      fractions. For the total, the different components have been
      added in quadrature.} \label{tab:2}
  \end{center}
\end{minipage}
\end{sideways}
\end{center}
\end{table}

The different decays used to model the right sign component are
collected into four categories:  $b\rightarrow
D_{(s)}l^-\bar\nu_l(X)$, $b\rightarrow D_{(s)}\pi,\rho,\omega,\dots$,
$b\rightarrow D_{(s)}D_s^-(X)$ and $b\rightarrow D_{(s)}\bar
D(X)$. To estimate the systematics related to the right sign
modelling, the relative contributions of $b\rightarrow
D_{(s)}l^-\bar\nu_l(X)$, $b\rightarrow D_{(s)}D_s^-(X)$ and
$b\rightarrow D_{(s)}\bar D(X)$ to the signal are varied. The ranges
are obtained from recent
measurements (Table~\ref{tab:2}). The weight of each
category is varied separately and, for the total, the different
contributions are added in quadrature.

Further contributions to the systematic error are: normalisation of
the $Z\rightarrow c\bar c$~background (uncertainty in $R_c$ and in
the fragmentation probabilities), normalisation of the combinatorial
background (uncertainty of the fit to the invariant mass distribution)
and uncertainty in $\epsilon_W/\epsilon_R$. 
For the $D^+\rightarrow
K^-\pi^+\pi^+$~sample which is particularly affected by the
combinatorial background, 
instead of using simulated data in the signal window,
the fit is repeated using real data candidates
selected outside the signal window. 
Both approaches are statistically consistent and the
difference in the fit result is interpreted as a systematic uncertainty
related to the combinatorial background shape.

\section{Conclusion}

The production of wrong sign charm mesons in $b$-hadron decay,
$b\rightarrow\bar D_{(s)}X$, $D_{(s)}=(D^0,D^+,D_s)$, was studied
using the DELPHI data collected in 1994 and 1995, leading to a
measurement of the fraction~${\mathcal
  B}(b\rightarrow\bar D_{(s)}X)/{\mathcal B}(b\rightarrow D_{(s)},\bar
D_{(s)}X)$. Combining this measurement with the branching
fractions~${\mathcal B}(b\rightarrow D^0,\bar D^0X)=(60.5\pm 3.2)\%$,
${\mathcal B}(b\rightarrow D^\pm X)=(23.7\pm 2.3)\%$ and ${\mathcal
  B}(b\rightarrow D_s^\pm X)=(18\pm 5)\%$~\cite{ref:10}, the following
result is obtained for wrong sign~$D$:
\begin{displaymath}
  {\mathcal B}(b\rightarrow\bar D^0X)+{\mathcal B}(b\rightarrow
  D^-X)=(9.3\pm 1.7(stat)\pm 1.3(syst)\pm 0.4({\mathcal B}))\% \; .
\end{displaymath}
The first uncertainty is statistical, the second one is the sum of all
systematic errors listed in Table~\ref{tab:2} (accounting for
correlated model systematics) and the last one
corresponds to the uncertainties in ${\mathcal B}(b\rightarrow D^0,\bar
D^0X)$ and ${\mathcal B}(b\rightarrow D^\pm X)$ (note that the quoted
statistical error includes both real data and Monte Carlo
statistics). This value is in good agreement with previous
measurements by CLEO~\cite{ref:2b} and ALEPH~\cite{ref:3}. The
inclusive branching fraction for wrong sign~$D_s$ is found to be:
\begin{displaymath}
  {\mathcal B}(b\rightarrow D_s^-X)=(10.1\pm 1.0(stat)\pm 0.6(syst)\pm
  2.8({\mathcal B}))\% \; .
\end{displaymath}
Again, the first uncertainty is statistical, the second one is the
total systematic error of Table~\ref{tab:2} and the last one
corresponds to the uncertainty on ${\mathcal B}(b\rightarrow D_s^\pm
X )$. This value agrees with the total $D_s$~production rate at
the $\Upsilon(4S)$, ${\mathcal B}(B\rightarrow D_s^\pm X)=(10.0\pm
2.5)\%$~\cite{ref:10}, where the dominant source of $D_s$~production is
double charm $b$-decay.

\subsection*{Acknowledgements}
\vskip 3 mm
 We are greatly indebted to our technical 
collaborators, to the members of the CERN-SL Division for the excellent 
performance of the LEP collider, and to the funding agencies for their
support in building and operating the DELPHI detector.\\
We acknowledge in particular the support of \\
Austrian Federal Ministry of Education, Science and Culture,
GZ 616.364/2-III/2a/98, \\
FNRS--FWO, Flanders Institute to encourage scientific and technological 
research in the industry (IWT), Belgium,  \\
FINEP, CNPq, CAPES, FUJB and FAPERJ, Brazil, \\
Czech Ministry of Industry and Trade, GA CR 202/99/1362,\\
Commission of the European Communities (DG XII), \\
Direction des Sciences de la Mati$\grave{\mbox{\rm e}}$re, CEA, France, \\
Bundesministerium f$\ddot{\mbox{\rm u}}$r Bildung, Wissenschaft, Forschung 
und Technologie, Germany,\\
General Secretariat for Research and Technology, Greece, \\
National Science Foundation (NWO) and Foundation for Research on Matter (FOM),
The Netherlands, \\
Norwegian Research Council,  \\
State Committee for Scientific Research, Poland, SPUB-M/CERN/PO3/DZ296/2000
and SPUB-M/CERN/PO3/DZ297/2000 \\
JNICT--Junta Nacional de Investiga\c{c}\~{a}o Cient\'{\i}fica 
e Tecnol$\acute{\mbox{\rm o}}$gica, Portugal, \\
Vedecka grantova agentura MS SR, Slovakia, Nr. 95/5195/134, \\
Ministry of Science and Technology of the Republic of Slovenia, \\
CICYT, Spain, AEN99-0950 and AEN99-0761,  \\
The Swedish Natural Science Research Council,      \\
Particle Physics and Astronomy Research Council, UK, \\
Department of Energy, USA, DE--FG02--94ER40817.

\end{document}